\documentclass[floatfix,showpacs,nofootinbib,twocolumn,showkeys]{revtex4-2}

\usepackage{graphicx}
\usepackage{etoolbox}
\usepackage{verbatim}
\usepackage{tensor}
\usepackage{url}
\usepackage{amsmath}
\usepackage{bm}
\usepackage{color}
\usepackage{hyperref}

\hypersetup{
    colorlinks=true,
    linkcolor=blue,
    filecolor=magenta,      
    urlcolor=blue,
    citecolor=blue
}

\usepackage{amsmath}
\usepackage{lineno}
\usepackage{xspace}

\usepackage{color}
\definecolor{orange}{rgb}{1,0.5,0}
\definecolor{darkgreen}{rgb}{0.0, 0.6, 0.22}

\begin{document}
%

\newcommand{\pp}           {pp\xspace}
\newcommand{\ppbar}        {\mbox{$\mathrm {p\overline{p}}$}\xspace}
\newcommand{\XeXe}         {\mbox{Xe--Xe}\xspace}
\newcommand{\PbPb}         {\mbox{Pb--Pb}\xspace}
\newcommand{\pA}           {\mbox{pA}\xspace}
\newcommand{\pPb}          {\mbox{p--Pb}\xspace}
\newcommand{\AuAu}         {\mbox{Au--Au}\xspace}
\newcommand{\dAu}          {\mbox{d--Au}\xspace}

\newcommand{\s}            {\ensuremath{\sqrt{s}}\xspace}
\newcommand{\snn}          {\ensuremath{\sqrt{s_{\mathrm{NN}}}}\xspace}
\newcommand{\pt}           {\ensuremath{p_{\rm T}}\xspace}
\newcommand{\meanpt}       {$\langle p_{\mathrm{T}}\rangle$\xspace}
\newcommand{\ycms}         {\ensuremath{y_{\rm CMS}}\xspace}
\newcommand{\ylab}         {\ensuremath{y_{\rm lab}}\xspace}
\newcommand{\etarange}[1]  {\mbox{$\left | \eta \right |~<~#1$}}
\newcommand{\yrange}[1]    {\mbox{$\left | y \right |~<~#1$}}
\newcommand{\dndy}         {\ensuremath{\mathrm{d}N_\mathrm{ch}/\mathrm{d}y}\xspace}
\newcommand{\dndeta}       {\ensuremath{\mathrm{d}N_\mathrm{ch}/\mathrm{d}\eta}\xspace}
\newcommand{\avdndeta}     {\ensuremath{\langle\dndeta\rangle}\xspace}
\newcommand{\dNdy}         {\ensuremath{\mathrm{d}N_\mathrm{ch}/\mathrm{d}y}\xspace}
\newcommand{\Npart}        {\ensuremath{N_\mathrm{part}}\xspace}
\newcommand{\Ncoll}        {\ensuremath{N_\mathrm{coll}}\xspace}
\newcommand{\dEdx}         {\ensuremath{\textrm{d}E/\textrm{d}x}\xspace}
\newcommand{\RpPb}         {\ensuremath{R_{\rm pPb}}\xspace}

\newcommand{\nineH}        {$\sqrt{s}~=~0.9$~Te\kern-.1emV\xspace}
\newcommand{\seven}        {$\sqrt{s}~=~7$~Te\kern-.1emV\xspace}
\newcommand{\twoH}         {$\sqrt{s}~=~0.2$~Te\kern-.1emV\xspace}
\newcommand{\twosevensix}  {$\sqrt{s}~=~2.76$~Te\kern-.1emV\xspace}
\newcommand{\five}         {$\sqrt{s}~=~5.02$~Te\kern-.1emV\xspace}
\newcommand{\twosevensixnn}{$\sqrt{s_{\mathrm{NN}}}~=~2.76$~Te\kern-.1emV\xspace}
\newcommand{\fivenn}       {$\sqrt{s_{\mathrm{NN}}}~=~5.02$~Te\kern-.1emV\xspace}
\newcommand{\LT}           {L{\'e}vy-Tsallis\xspace}
\newcommand{\GeVc}         {Ge\kern-.1emV/$c$\xspace}
\newcommand{\MeVc}         {Me\kern-.1emV/$c$\xspace}
\newcommand{\TeV}          {Te\kern-.1emV\xspace}
\newcommand{\GeV}          {Ge\kern-.1emV\xspace}
\newcommand{\MeV}          {Me\kern-.1emV\xspace}
\newcommand{\GeVmass}      {Ge\kern-.2emV/$c^2$\xspace}
\newcommand{\MeVmass}      {Me\kern-.2emV/$c^2$\xspace}
\newcommand{\lumi}         {\ensuremath{\mathcal{L}}\xspace}

\newcommand{\ITS}          {\rm{ITS}\xspace}
\newcommand{\TOF}          {\rm{TOF}\xspace}
\newcommand{\ZDC}          {\rm{ZDC}\xspace}
\newcommand{\ZDCs}         {\rm{ZDCs}\xspace}
\newcommand{\ZNA}          {\rm{ZNA}\xspace}
\newcommand{\ZNC}          {\rm{ZNC}\xspace}
\newcommand{\SPD}          {\rm{SPD}\xspace}
\newcommand{\SDD}          {\rm{SDD}\xspace}
\newcommand{\SSD}          {\rm{SSD}\xspace}
\newcommand{\TPC}          {\rm{TPC}\xspace}
\newcommand{\TRD}          {\rm{TRD}\xspace}
\newcommand{\VZERO}        {\rm{V0}\xspace}
\newcommand{\VZEROA}       {\rm{V0A}\xspace}
\newcommand{\VZEROC}       {\rm{V0C}\xspace}
\newcommand{\Vdecay} 	   {\ensuremath{V^{0}}\xspace}

\newcommand{\ee}           {\ensuremath{e^{+}e^{-}}} 
\newcommand{\pip}          {\ensuremath{\pi^{+}}\xspace}
\newcommand{\pim}          {\ensuremath{\pi^{-}}\xspace}
\newcommand{\kap}          {\ensuremath{\rm{K}^{+}}\xspace}
\newcommand{\kam}          {\ensuremath{\rm{K}^{-}}\xspace}
\newcommand{\pbar}         {\ensuremath{\rm\overline{p}}\xspace}
\newcommand{\kzero}        {\ensuremath{{\rm K}^{0}_{\rm{S}}}\xspace}
\newcommand{\lmb}          {\ensuremath{\Lambda}\xspace}
\newcommand{\almb}         {\ensuremath{\overline{\Lambda}}\xspace}
\newcommand{\Om}           {\ensuremath{\Omega^-}\xspace}
\newcommand{\Mo}           {\ensuremath{\overline{\Omega}^+}\xspace}
\newcommand{\X}            {\ensuremath{\Xi^-}\xspace}
\newcommand{\Ix}           {\ensuremath{\overline{\Xi}^+}\xspace}
\newcommand{\Xis}          {\ensuremath{\Xi^{\pm}}\xspace}
\newcommand{\Oms}          {\ensuremath{\Omega^{\pm}}\xspace}
\newcommand{\degree}       {\ensuremath{^{\rm o}}\xspace}

\newcommand{\gab}          {\ensuremath{\langle \cos(\varphi_\alpha + \varphi_\beta - 2\psid) \rangle}\xspace}

\newcommand{\dab}          {\ensuremath{\langle \cos(\varphi_\alpha - \varphi_\beta) \rangle}\xspace}

\newcommand{\gc}           {\ensuremath{\gamma_{\alpha\beta}}\xspace}

\newcommand{\dc}         {\ensuremath{\delta_{\alpha\beta}}\xspace}
\newcommand{\psid}         {\ensuremath{\Psi_{\rm 2}}\xspace}

\newcommand{\fcme}         {\ensuremath{f_{\rm CME}}\xspace}
\newcommand{\fcmexe}         {\ensuremath{f_{\rm CME}^{\rm Xe-Xe}}\xspace}
\newcommand{\fcmepb}         {\ensuremath{f_{\rm CME}^{\rm Pb-Pb}}\xspace}

\newcommand{\todo}[1]{\textcolor{darkgreen}{[TODO: \emph{#1}]}}
\newcommand{\done}[1]{\textcolor{blue}{[DONE: \emph{#1}]}}
\newcommand{\new}[1]{\textcolor{red}{#1}}
\newcommand{\old}[1]{\textcolor{orange}{#1}}
\newcommand{\change}[1]{\textcolor{blue}{[CHANGE: #1]}}
\newcommand{\ask}[1]{\textcolor{magenta}{\bf [ #1]}}

\title{Probing the magnetic field strength dependence of the Chiral Magnetic Effect}
\date{\today}

\author{Panos~Christakoglou}
\affiliation{Nikhef, Amsterdam, 1098 XG, The Netherlands}

\begin{abstract}
The article presents a study aimed at probing the dependence of the Chiral Magnetic Effect (CME) on the magnetic field strength using the Anomalous Viscous Fluid Dynamics (AVFD) model in Pb--Pb at LHC energies. The results demonstrate the quadratic dependence of the correlators used for the study of the CME in heavy ion collisions on the number of spectators, a proxy of the magnitude of the magnetic field. The article also presents the  extension of this approach to a two dimensional space, formed by both the aforementioned proxy of the magnetic field strength but also a proxy of the final state ellipticity, a key ingredient of the background in these measurements, for each centrality interval. This provides an exciting possibility to experiments to isolate the background contributions from the potential CME signal.
\end{abstract}

\maketitle

\section{Introduction}
\label{Sec:Introduction}

The chiral magnetic effect (CME)~\cite{Fukushima:2008xe} is the development of an electric current $\vec{J}$, induced by a chirality imbalance between left- and right-handed chiral fermions characterised by a chiral chemical potential $\mu_5$, that develops parallel to an external magnetic field ($\vec{B}$) according to 

\begin{center}
    \begin{equation}
        \vec{J} = \sigma_5 \vec{B}.
    \end{equation}
\end{center}
    
\noindent In the equation above $\sigma_5$ is the chiral magnetic conductivity that is proportional to $\mu_5$. This chirality imbalance in theories like quantum chromodynamics or QCD is connected to transitions between different vacuum states of the theory and is, consequently, a reflection of fundamental symmetries such as parity (P) and its combination with charge conjugation (C) being broken~\cite{Lee:1973iz,Lee:1974ma,Morley:1983wr}. An exiting possibility emerged under the realisation that such effects can be accessed experimentally using heavy ion collisions accelerated at ultrarelativistic energies like the ones achieved at the Relativistic Heavy Ion Collider (RHIC) or the Large Hadron Collider (LHC)~\cite{Kharzeev:1998kz,Kharzeev:1999cz,Kharzeev:2015kna,Kharzeev:2007tn,Kharzeev:2007jp,Kharzeev:2015znc,Li:2020dwr,Kharzeev:2020jxw,Zhao:2019hta}. These collisions can create extreme conditions of energy density and temperature which exceed the necessary values expected by lattice-QCD calculations~\cite{Bazavov:2009zn,Bazavov:2011nk,Borsanyi:2010cj} to reach a state of matter called Quark Gluon Plasma (QGP)~\cite{Karsch:2003jg} which consists of strongly coupled chiral fermions and gluons~\cite{PHOBOS:2004zne,STAR:2005gfr,PHENIX:2004vcz,BRAHMS:2004adc,ALICE:2022wpn}. In addition, in non-central heavy ion collisions i.e. in collisions with large values of impact parameter, the charged nucleons that do not reside in the overlap region also called ``spectators'', fly away with large velocities and can generate large values of magnetic fields. This magnetic field, that can reach magnitudes larger than $10^{16}$~T~\cite{Skokov:2009qp,Bzdak:2011yy,Deng:2012pc,Christakoglou:2021nhe} at the LHC,  decays rapidly with a rate that depends on the electrical conductivity of the QGP, a property of the medium which is unconstrained experimentally.

The search for the discovery of the CME intensified after Voloshin in Ref.~\cite{Voloshin:2004vk} proposed a sensitive experimental observable that relies on measuring two-particle azimuthal correlations relative to the reaction plane ($\Psi_{\mathrm{RP}}$), the plane defined by the impact parameter and the beam axis, according to 

\begin{center}
    \begin{equation}
        \gamma = \langle \cos(\varphi_{\alpha} + \varphi_{\beta} - 2\Psi_{\rm RP}) \rangle,
    \end{equation}
    \label{Eq:Gamma}
\end{center}

\noindent where $\alpha$ and $\beta$ indicate particles with the same or opposite charge. This expression can probe the first coefficient $a_1$\footnote{A way to probe the CME is by introducing coefficients $a_{n}$ in the Fourier series frequently used in  studies of azimuthal anisotropy~\cite{Voloshin:1994mz}. This leads to the expression $\frac{dN}{d\varphi} \approx 1 + 2\sum_{n} \Big[v_n \cos[n(\varphi - \Psi_n)] + a_n \sin[n(\varphi - \Psi_n)]\Big]$, where $N$ is the number of particles, $\varphi$ is the azimuthal angle of the particle and $v_n$ are the corresponding flow coefficients ($v_1$: directed flow, $v_2$: elliptic flow, $v_3$: triangular flow etc.). The n-th order symmetry plane of the system, $\Psi_n$, is introduced to take into account that the overlap region of the colliding nuclei exhibits an irregularshape~\cite{Manly:2005zy,Bhalerao:2006tp,Alver:2008zza,Alver:2010gr,Alver:2010dn}. } which quantifies the magnitude of the CME signal and more specifically it is proportional to correlations between the leading terms for different charge combinations $\langle a_{1,\alpha} a_{1,\beta} \rangle$. In parallel, one can also measure the two particle correlator that has no dependence on the reaction plane, of the form

\begin{center}
    \begin{equation}
        \delta = \langle \cos(\varphi_{\alpha} - \varphi_{\beta}) \rangle,
    \end{equation}
    \label{Eq:Delta}
\end{center}

\noindent This correlator is still sensitive to the potential signal from $\langle a_{1,\alpha} a_{1,\beta} \rangle$ but is dominated by background contributions as discussed and demonstrated in Ref.~\cite{Christakoglou:2021nhe}. Consequently, the article does not focus on or presents the corresponding results.

The first experimental measurements using this approach were reported by the STAR Collaboration in Au--Au collisions at $\sqrt{s_{\mathrm{NN}}} = 0.2$~TeV~\cite{Abelev:2009ac,Abelev:2009ad} and were consistent with initial expectations for a charge separation relative to the reaction plane due to the CME. Since then, many more attempts not only at RHIC~\cite{STAR:2013ksd,STAR:2014uiw,STAR:2013zgu,STAR:2019xzd,STAR:2020gky,STAR:2021pwb} but also at the LHC~\cite{Abelev:2012pa,Acharya:2020rlz,Khachatryan:2016got,Sirunyan:2017quh} reported results that were not able, at least until this point, to identify unambiguously the existence of the CME. One of the main reasons is the fact that these measurements are dominated by background sources~\cite{Wang:2009kd,Bzdak:2009fc,Liao:2010nv,Schlichting:2010qia}, the most prominent of which is the combination of local charge conservation or LCC (i.e. the production of oppositely charged particles from a neutral fluid element) in combination with how the QGP expands outwards in an anisotropic way. This latter is encapsulated by the phenomenon of anisotropic flow, which is usually quantified by the anisotropic flow coefficients $v_n$ of the Fourier expansion of the azimuthal particle distribution in the final state of a heavy ion collision. 

After these first results the field moved in two parallel directions: the first one focuses in constraining and quantifying the background while in the second one modifies some of the components of the signal and looks for relative changes in the measurement. A characteristic example of the first direction is the event shape engineering (ESE) studies that allows to select events with different magnitude of ellipticity within the same centrality~\cite{Schukraft:2012ah}. This allows to quantify the dependence of the measured charge dependent differences, $\Delta \gamma$, on one of the main background components i.e. $v_2$ or elliptic flow, the second and most dominant flow coefficient. On the other side of the spectrum, the STAR collaboration used two isobar systems~\cite{STAR:2021mii}, namely $^{96}_{40}Zr-^{96}_{40}Zr$ and $^{96}_{44}Ru-^{96}_{44}Ru$, that are very similar in size and thus the background contributions to the measurements were expected to be very similar. At the same time, however, the $Ru$-nucleus contains 10\% more protons than the $Zr$ which results in a significantly larger magnitude of $\vec{B}$. This, consequently, is expected to result in a larger contribution to $\Delta \gamma$ originating from the CME signal, if any, in $Ru-Ru$ than in $Zr-Zr$ collisions. In both cases, either using the ESE or different isobars, the results are consistent with no CME contribution and led to the extraction of upper limits~\cite{STAR:2020gky,Sirunyan:2017quh,Acharya:2017fau}. 

The study presented in this article explores the dependence of $\Delta \gamma$ on the magnitude of the magnetic field. In particular, since the first order coefficient $a_1$ is proportional to the value of $\mu_5$ but also to the magnitude of $\vec{B}$, the correlator of Eq.~\ref{Eq:Gamma} is expected to have a quadratic dependence on both quantities. Within models the value of the magnetic field can be calculated for each centrality interval and can be connected to the number of spectator nucleons. Furthermore, since within a given centrality interval this latter number is expected to fluctuate from event to event, one can devise a strategy for selecting events with large or small number of spectators which would, consequently, result into large or small magnitude of $\vec{B}$. The combination of this with the ESE technique could provide a powerful tool to isolate or fix the background contribution while probing in parallel the quadratic dependence of $\Delta \gamma$ on $\vec{B}$. Experimentally, triggering on events with different values of spectators within a given centrality can be realised with the energy deposited by these non-interacting nucleons on zero-degree calorimeters that are positioned close to the beam pipe far away from the interaction region.

This article presents the strategy for such study in Pb--Pb collisions at $\sqrt{s_{\mathrm{NN}}} = 5.02$~TeV with the Anomalous-Viscous Fluid Dynamics (AVFD) framework~\cite{Shi:2017cpu,Jiang:2016wve,Shi:2019wzi}. The next section discusses some details about the model, the sample that was analysed and illustrates the connection between the magnitude of $\vec{B}$ and the number of spectators as well as the ESE procedure. Section~\ref{Sec:Results} presents the main results and is followed by the summary.
\section{Model and analysis details}
\label{Sec:Model}
The study was performed over a sample of Pb--Pb collisions at $\sqrt{s_{\mathrm{NN}}} = 5.02$~TeV generated with the AVFD model. This state-of-the-art model describes the initial state of the collision using a Glauber prescription, and accounts for the development of the early stage electromagnetic fields as well as for the propagation of anomalous fermion currents. The expanding medium is treated after 0.6~fm/$c$ with a 2+1 dimensional viscous hydrodynamics (VISH2+1) code using values of shear and bulk viscosities over entropy density of $\eta/s$ = 0.08 and $\zeta/s = 0$. Beyond a decoupling energy density of $\epsilon = 0.18$~fm$/c^3$ the system is described by a hadron cascade model (UrQMD)~\cite{Bass:1998ca}. In addition, the model allows for the inclusion of a non-zero axial current density $n_5/\mathrm{s}$ which dictates the imbalance between right- and left-handed fermions induced in the initial stage of each event. This, consequently, leads to a CME signal in the final state. Furthermore the background contribution in this measurement is controled by the percentage of positive and negative charged partners emitted from the same fluid element relative to the total multiplicity of the event, referred to from now on as LCC percentage. Both values of $n_5/\mathrm{s}$ and LCC percentage are identical to the ones reported in Ref.~\cite{Christakoglou:2021nhe} where the model was tuned to describe the experimental measurements reported at LHC energies. 

Around 100K events were produced for the centrality interval 10--70\% defined by different impact parameter ranges, in steps of 10\%. An additional sample of 1M events for the 50--60\% centrality interval was generated to allow for the extension of the analysis using the ESE method. The centrality interval 0--10\% is not studied in this article based on the expectation that the magnetic field is significantly smaller in central than in semi-central and peripheral Pb--Pb collisions, leading to a smaller CME signal. An additional, technical reason for not studying this centrality interval is related to increased requirements for computing resources. The analysis is performed in the same kinematic ranges as the experimental measurements for primary charged particles that are emitted within a pseudorapidity of $|\eta| < 0.8$ and have transverse momentum of $0.2 < p_{\mathrm{T}} < 5$~GeV/$c$. 

\begin{figure}[!h]
    \begin{center}
    \includegraphics[width = 0.5\textwidth]{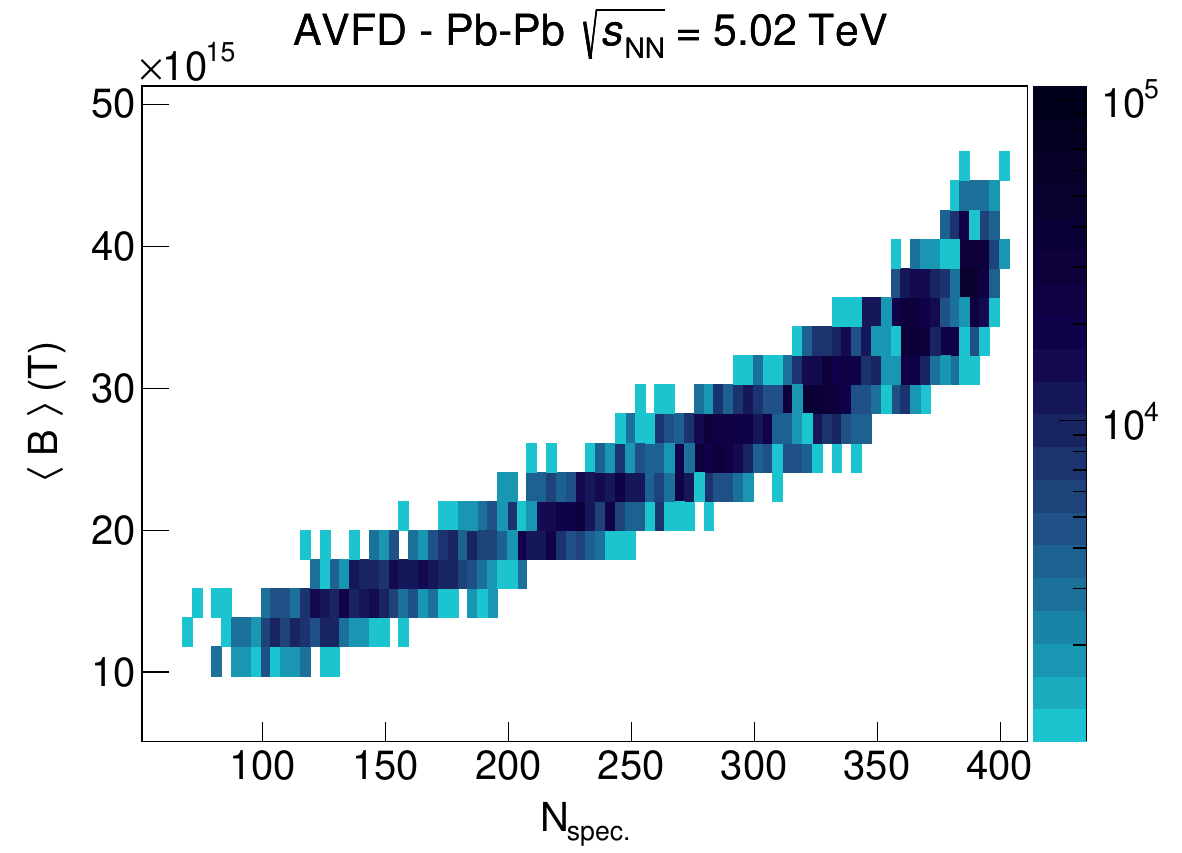}
    \end{center}
    \caption{The correlations between the number of spectators and the magnitude of the magnetic field generated in Pb--Pb collisions at $\sqrt{s_{\mathrm{NN}}} = 5.02$~TeV as
    obtained from the tuned AVFD sample (see text for details).}
    \label{fig:NspecVsB}
\end{figure}

The model gives in addition the possibility to calculate but also evolve the value of the magnetic field that decays with time according to

\begin{equation}
    B(\tau,x)=\frac{B_0}{1+\tau^2/\tau_B^2},
\end{equation}

\noindent where $\tau_B$ is the magnetic field lifetime which is set, in this work, 
conservatively to 0.2~fm/$c$, similarly to what was done in Ref.~\cite{Christakoglou:2021nhe}. At the same time and for each centrality interval one can calculate the number of spectator nucleons, $\mathrm{N_{spec.}}$, from the Glauber model. Figure~\ref{fig:NspecVsB} presents the dependence of the magnitude of the magnetic field on $\mathrm{N_{spec.}}$, where a clear correlation can be observed. From this plot it also becomes evident that for each centrality interval the number of spectators and thus the magnitude of the magnetic field fluctuates from event to event. One can thus define percentiles from the distribution of the number of spectators e.g. 25\% highest or lowest number of spectators and map these events to events where the magnetic field is largest or smallest within a given centrality interval. In this work, every centrality interval is split in four subsamples that correspond to different percentiles of number of spectators, from 0\% to 100\% with a step of 25\%.

In parallel, one can also trigger on events where one of the main components that drives the background, the elliptic flow or $v_2$, is small or large within the same centrality. This is done by calculating the magnitude of the second-order reduced flow vector, $q_2$, which is defined according to 

\begin{equation}
    q_2 = \frac{|Q_2|}{\sqrt{M}},
\end{equation}

\noindent where $Q_2 = \sqrt{Q_{2,x}^2 + Q_{2,y}^2}$ is the magnitude of the second order harmonic flow vector and $M$ is the multiplicity. In this work, the vector $Q_2$ is calculated from the azimuthal distribution of primary charged particles emitted in the pseudorapidity range $-3.7 < \eta < -1.7$, thus simulating the acceptance of one of the scintillator counters of ALICE at the LHC, namely the V0C detector~\cite{ALICE:2008ngc}.

\begin{figure}[!h]
    \begin{center}
    \includegraphics[width = 0.5\textwidth]{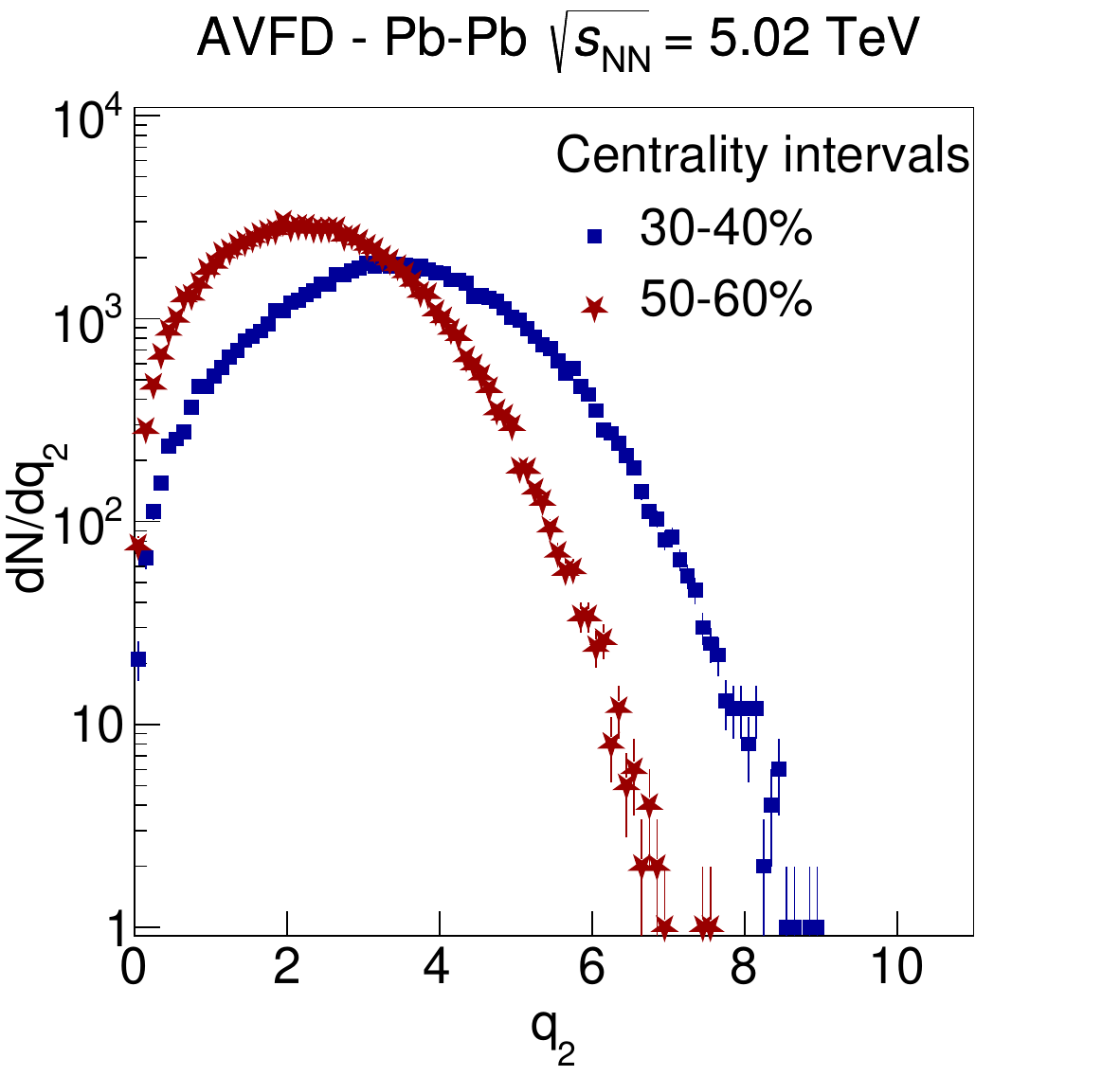}
    \end{center}
    \caption{The distribution of $q_2$ for two different centrality intervals of Pb--Pb collisions at $\sqrt{s_{\mathrm{NN}}} = 5.02$~TeV as obtained from the tuned AVFD sample (see text for details).}
    \label{fig:Q2Distributions}
\end{figure}

Figure~\ref{fig:Q2Distributions} presents the $q_2$ distributions for two indicative centrality intervals, i.e. 30-40\% and 50-60\%, of Pb--Pb collisions at $\sqrt{s_{\mathrm{NN}}} = 5.02$~TeV. These distributions are then split in the relevant percentiles with a 25\% step, thus alowing the selection of events that are characterised by different magnitude of final state ellipticity, ranging from the 25\% lowest to the 25\% highest $q_2$ values. 

The combination of triggering on events with different value of number of spectators and thus magnetic filed and, at the same time, different value of $q_2$ within a given centrality interval forms a two-dimensional space where the contribution from the signal and background components, respectively, can be varied and controlled in an efficient way. Considering also the expectation for a quadratic dependence of $\Delta \gamma$ on the value of $\vec{B}$ and the relevant scaling that the same observable has with $v_2$, the strategy creates a powerful tool that could also be used experimentally with clear and unique expectations from theory that can also be demonstrated using suitable models.
\section{Results}
\label{Sec:Results}

\begin{figure}[!h]
    \begin{center}
    \includegraphics[width = 0.5\textwidth]{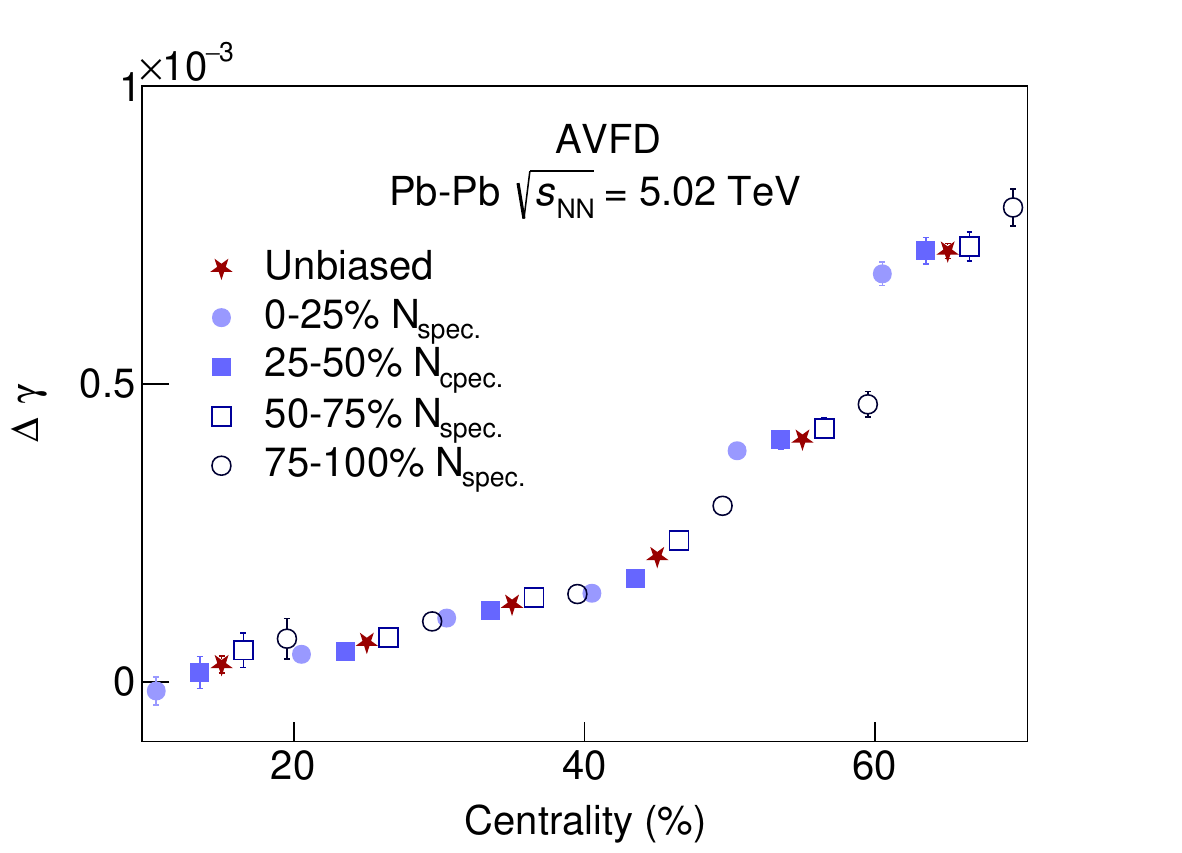}
    \end{center}
    \caption{The centrality dependence of $\Delta \gamma$ for various percentiles of $\mathrm{N_{spec.}}$. The results are obtained from the analysis of Pb--Pb collisions at $\sqrt{s_{\mathrm{NN}}} = 5.02$~TeV produced with the AVFD model which is tuned to describe the experimental measurements (see text for details).}
    \label{fig:CentralityDependence}
\end{figure}

Figure~\ref{fig:CentralityDependence} presents the centrality dependence of $\Delta \gamma$ and 
as obtained from the analysis of the AVFD samples of Pb--Pb collisions at $\sqrt{s_{\mathrm{NN}}} = 5.02$~TeV. The red filled star markers correspond to the unbiased results and are compatible with the ones presented in Refs.~\cite{Christakoglou:2021nhe,ALICE:2022wpn} where it was shown that they describe at the same time both observables. For each centrality interval, this data point is accompanied with the corresponding results for different selections in the number of spectators from the lowest 25\% to the highest 25\% of the $\mathrm{N_{spec.}}$ distribution, with a 25\% step. It can be seen that the magnitude of $\Delta \gamma$
increases with increasing $\mathrm{N_{spec.}}$ and, consequently as discussed in Section~\ref{Sec:Model}, with the magnitude of the magnetic field. This increase seems to be of quadratic nature, as expected from the dependence of $\Delta \gamma$ on the value of $\vec{B}$. Similar observations can be made for the rest of the centrality intervals (not shown in this article).

\begin{figure}[!h]
    \begin{center}
    \includegraphics[width = 0.5\textwidth]{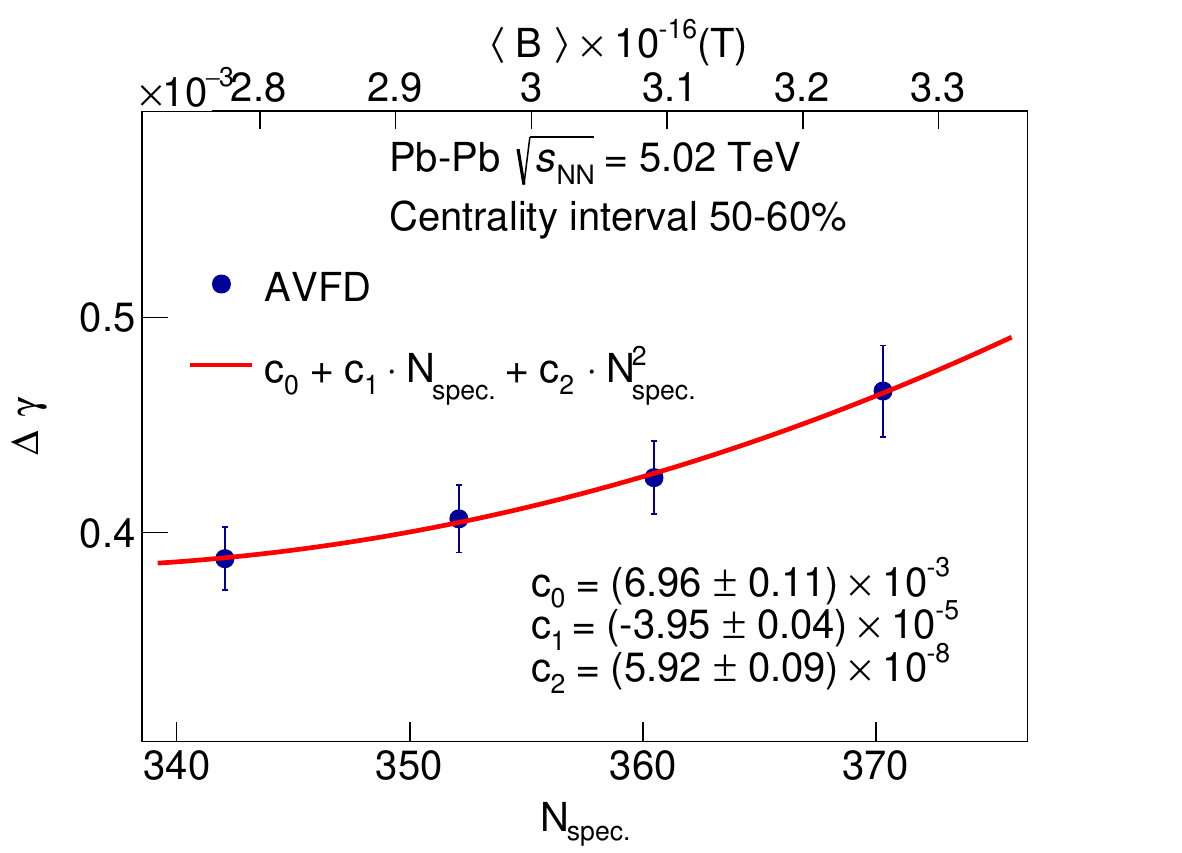}
    \end{center}
    \caption{The dependence of $\Delta \gamma$ on the number of spectators for the 50--60\% centrality interval of Pb--Pb collisions at $\sqrt{s_{\mathrm{NN}}} = 5.02$~TeV. The data points are fitted with a second order polynomial represented by the red solid line.}
    \label{fig:CorrelatorsVsSpectators}
\end{figure}

In order to further illustrate but also quantify this behavior, figure~\ref{fig:CorrelatorsVsSpectators} presents the dependence of $\Delta \gamma$ 
on $\mathrm{N_{spec.}}$ for one indicative centrality interval i.e., 50--60\% Pb--Pb collisions at $\sqrt{s_{\mathrm{NN}}} = 5.02$~TeV. The data points are fitted with a second order polynomial that yields a significantly non-zero value of the second order coefficient. This unambiguously confirms the expected behavior that rises from the dependence of the correlator on the term $\langle a_{1,\alpha} a_{1,\beta} \rangle$. This, in turns, gives rise to the quadratic dependence of $\Delta \gamma$ on the value of $\vec{B}$.

\begin{figure}[!h]
    \begin{center}
    \includegraphics[width = 0.5\textwidth]{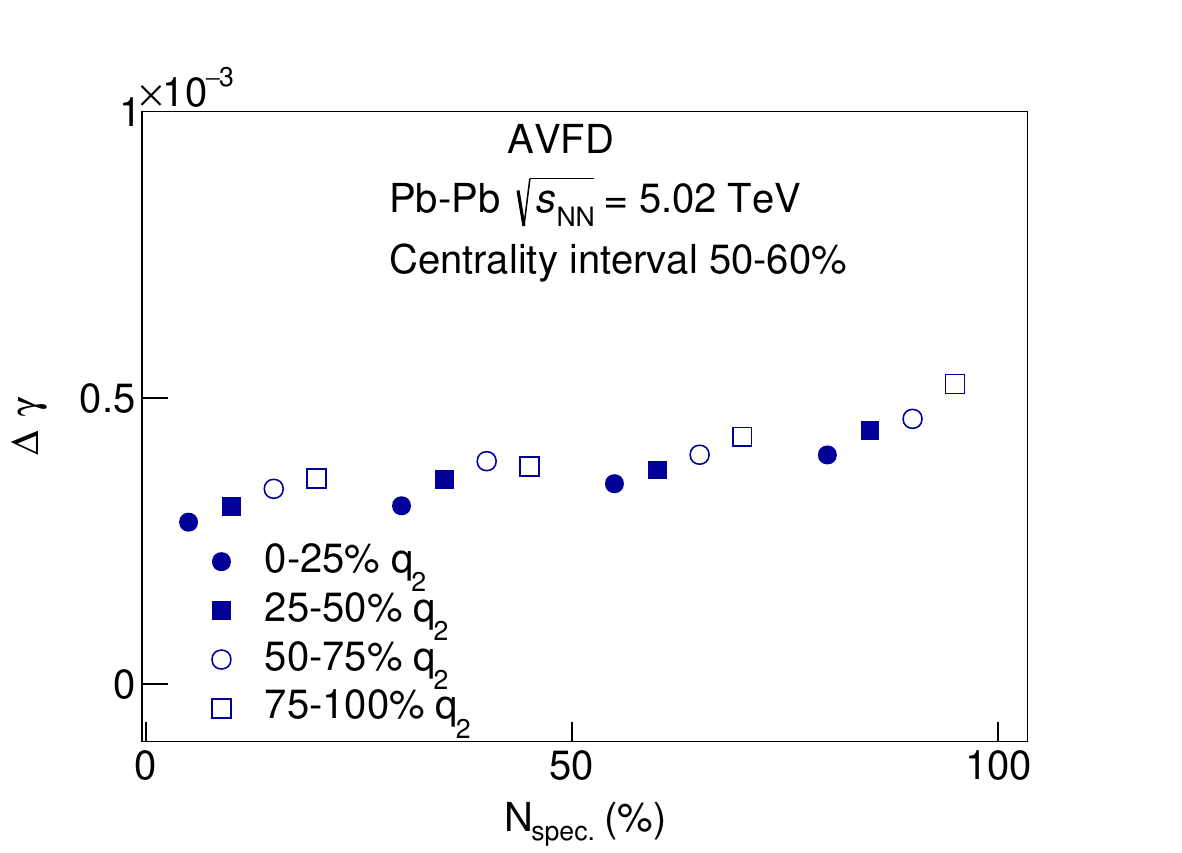}
    \end{center}
    \caption{The dependence of $\Delta \gamma$ on the number of spectators expressed in qualtiles of the relevant distribution for the 50-60\% centrality interval of Pb--Pb collisions at $\sqrt{s_{\mathrm{NN}}} = 5.02$~TeV. For each $\mathrm{N_{spec.}}$ percentile interval, results for different $q_2$ percentiles are shown.}
    \label{fig:CorrelatorsVsQ2}
\end{figure}

Finally, figure~\ref{fig:CorrelatorsVsQ2} presents $\Delta \gamma$ 
for events with different percentiles of the distribution of the number of spectators for the 50-60\% centrality interval of Pb--Pb collisions at $\sqrt{s_{\mathrm{NN}}} = 5.02$~TeV. Each percentile range of $\mathrm{N_{spec.}}$ contains results for the various $q_2$ selections, ranging from the 25\% lowest to the 25\% highest $q_2$ values, with a step of 25\%. The linear dependence of $\Delta \gamma$  on $q_2$ and, consequently, on $v_2$ (a major component of the background contribution) is evident. In parallel, the values of $\Delta \gamma$ increase quadratically as a function of the $\mathrm{N_{spec.}}$ for a fixed $q_2$ interval. This was verified by fitting the corresponding dependence of the data points with a second order polynomial, similar to what is presented in fig.~\ref{fig:CorrelatorsVsSpectators}. 
This two dimensional grid formed by the proxies of the magnitude of the magnetic field and the final state ellipticity provides a powerful tool in experiments to disentangle the dominating background contributions in the measurements from the potential CME signal.

\section{Summary}
\label{Sec:Summary}
In this article, a new way of probing the magnetic field dependence of the Chiral Magnetic Effect is presented using the Anomalous-Viscous Fluid Dynamics framework~\cite{Shi:2017cpu,Jiang:2016wve}. The two correlators used regularly in the search of the CME i.e., $\Delta \gamma$ and $\Delta \delta$, were used to analyse samples of AVFD generated Pb--Pb collisions at $\sqrt{s_{\mathrm{NN}}} = 5.02$~TeV. The results demonstrated a quadratic dependence of both correlators with increasing number of spectators, the latter being a proxy for the magnitude of the early stage magnetic field. Finally, the extension of this study to a two dimensional space, formed by the number of spectators and a proxy of the final state ellipticity for each centrality interval, provides an exciting possibility to isolate experimentally the contributions of the background and the potential CME signal. 

\begin{acknowledgements}
I am grateful to Prof.~Jinfeng Liao and Dr.~Shuzhe Shi for providing the source code of the model, for their guidance and their feedback during this study. I would also like to thank Prof. Sergei Voloshin for his suggestions but also Prof. Dima Kharzeev for the enlightening discussion. I am also thankful and acknowledge the stimulating discussions with members of my group such as Shi Qiu and Jasper Westbroek.

\end{acknowledgements}

\bibliographystyle{utphys}
\bibliography{cmeZDCPaperRefs}{}

\end{document}